\documentclass{article}

\author{}

\usepackage{amsmath}
\usepackage{amsthm}
\usepackage{amsfonts}
\usepackage{amssymb}
\usepackage{latexsym}
\usepackage{setspace}
\usepackage[dvips]{graphicx}
\usepackage{pstricks}

\newcommand{\bc}{\begin{center}}
\newcommand{\ec}{\end{center}}
\newcommand{\be}{\begin{equation}}
\newcommand{\ee}{\end{equation}}
\newcommand{\bq}{\begin{eqnarray}}
\newcommand{\eq}{\end{eqnarray}}
\newcommand{\nn}{\nonumber}

\newcommand{\ed}{\end{document}}

\begin{document}

\title{Betting Around the Clock \\ 
Time Change and Long Term Model Risk\thanks{This paper was suggested to me by an interesting debate and questions with Gabriel Pereira Pinto and Stephen Mittenhall on my previous work with Peter Carr. It is part of a project that has been funded by European Union – Next Generation EU, Mission 4 Component 2, CUP J53D23004180006. The author is also grateful to Giulio Bottazzi, Pasquale Cirillo, Pietro Dindo, Dilip Madan, Filippo Massari, Lorenzo Torricelli for useful comments. Remaining errors are my own.  } 
}
\author{ Umberto Cherubini \footnote{University of Bologna, Department of Economics, Piazza Scaravilli 2, 40126 Bologna, e-mail umberto.cherubini@unibo.it} } 
\date{}
\maketitle
\center{\small{This version: September 12th 2025}}
\abstract{\noindent 	We investigate the performance of the Kelly rule in a setting in which the dynamics of the return is represented by a time change process. We find that in this general semi-martingale setting the Kelly rule does not maximize the average growth rate, unless the log-return is normally distributed. Namely, the investment position proposed by the Kelly rule is too large, and the investor could achieve a higher average growth rate by investing less aggressively. The higher the variance of the stochastic clock, the more material the failure of the Kelly rule. The ruin threshold proposed by Thorp (1969) is closer, even though examples based on stochastic clock variance estimates taken from the literature show that Kelly rule investment remains safely in the ruin-free region. Finally, the goal of keeping the investment below the ruin threshold for a family of stochastic clock distributions generates a long term investment problem that parallels the \lq \lq acceptable investment" theory.        }
\pagebreak
\doublespace

\section{Introduction}
Kelly rule has become a consolidated, albeit lively debated, result in investment science. The main idea is that all that matters for long term decisions with reinvestment of capital in favorable games is the law of large numbers. Under the maintained hypothesis of geometric compounding the law of large numbers apply to the logarithm of gross returns, and finally the average long term return would converge to its expected value. So, the best an investor could do is maximizing the log-return period by period and waiting for the law of large numbers to work. The principle was developed by Kelly (1956), and rigorously proved by Breiman (1961) and Thorp (1969). The theory attracted huge criticism from economists, that opposed the paramount role of risk aversion over the law of large numbers. The harshest reply came from Samuelson (1963,1971), while Markowitz (1976) recognized that the Kelly criterion could not be beaten in the long run. 

The maintained hypothesis throughout this controversy was that the returns are i.i.d. Taking a broad view of the processes with i.i.d. increments, Carr and Cherubini (2022) proved that the Kelly rule requires the more restrictive assumption that gross returns be lognormally distributed. The proof exploits the fundamental result by Monroe (1978) that shows that all the semimartingale processes can be represented as arithmetic brownian motion processes stopped at random times. The law of the stochastic clock fully determines the dynamics of the log-returns. Carr and Cherubini (2022) show that when the law of the stochastic clock is brought into the picture the geometric average return is actually a Kolmogorov average. In this average the exponential function of the geometric average is substituted by the moment generating function of the stochastic clock in the unit time. Only if the time is not stochastic, and the dynamics of wealth is driven by a geometric brownian motion, the law of large numbers would apply to the logarithm, as in the original Kelly result. The consequence of this finding is that for most of the models that are currently applied to the dynamics of asset prices, the Kelly rule of maximization is not optimal, even if the returns are assumed to be i.i.d. Actually, it is only optimal if gross returns are log-normal. 

It is of course very difficult to convince the audience that a consolidated result as the Kelly rule holds only under the much more restrictive assumption of normally distributed returns, that has been dismissed for long time. Long standing concepts like Growth-Optimal-Portfolios have been built on the concept of log-wealth maximization. Can we dare to affirm that this is \lq \lq growth optimal" only if gross returns are log-nomally distributed? Here we take on a different strategy. We ask what happens if a long term investor applies the Kelly rule to a bet that is not log-normal: for example, what happens if one uses the Kelly rule when the dynamics of returns is Variance Gamma, one of the most widely used models in asset pricing in substitution of the log-normal distribution? Can he actually reach the average growth predicted by the Kelly rule? Could he do any better by selecting a different rule? 

We find answers to these questions remaking a paper by Rotando and Thorpe (1992) in which they ask the same questions about the Kelly criterion. They ask: what happens if one does not follow the Kelly criterion? We ask the question: what happens if one uses the Kelly criterion but the dynamics of returns is, for example, Variance Gamma? 

To fix the notation, the problem is to maximize the average return on 
\begin{equation}\nonumber
\frac{W_N}{W_0}= R_1^{\Delta t}\times R_2^{\Delta t} \times \ldots \times R_{N-1}^{\Delta t} \times R_N^{\Delta t} 
\end{equation}
which may be written as
\begin{equation}\nonumber
\frac{W_N}{W_0}= \exp(\log(R_1) {\Delta t} + \log(R_2) {\Delta t} +\ldots \log(+R_{N-1}){\Delta t}+\log(R_N){\Delta t})  
\end{equation}
Carr and Cherubini (2022) analyze the general (semi-martingale) case in which gross returns $R_i$ follow a subordinated process. They assume
\begin{equation}\nonumber
R_i = \exp(s_i Z_i \Delta t)
\end{equation} 
where $s_i$ are i.i.d normal random variables and $Z_i$ is a subordinator, that is sequence of strictly positive i.i.d random variables with $E(Z_i)= 1$. In this case, is we set the time period $\Delta t = 1$ for simplicity, the geometric average of the returns can be obtained as an average of the returns across different scenarios of the stochastic clock as
\begin{equation}\label{main}
E\left(\frac{W_N}{W_0}\right)^{\frac{1}{N}} =\psi \left(\frac{\psi^{-1}(\tilde R_1) + \psi^{-1}( \tilde R_2) +\ldots + \psi^{-1}(\tilde R_{N-1})+\psi^{-1}(\tilde R_N)}{N} \right)
\end{equation}         
where $\psi(\cdot)$ is the moment generating function of the stochastic clock $Z$ and we define
\begin{equation}\label{semi_martingale_return}
\tilde R_i \equiv E\left(e^{s_i Z_i}\right) = \psi (s_i)
\end{equation}  
Notice that the meaning of the expectation $E(\cdot)$ in equations \ref{main} and \ref{semi_martingale_return} is that the geometric mean return, and every one pediod return, are averaged across the scenarios of the stochastic clock. Carr and Cherubini (2022) observe that only if the distribution of the clock is degenerate, with $Z_i =1,\forall i$ we have
\begin{equation}\nonumber
 \frac{W_N}{W_0}^{\frac{1}{N}} =\exp \left(\frac{\log( R_1) + \log( R_2) +\ldots + \log( R_{N-1})+\log(R_N)}{N} \right)
\end{equation}         
since obviously $ \tilde R_i = R_i = e^{s_i}, \forall i$. So, the concluding remark by Kelly (1956),  
\lq \lq \textit{... that it is the logarithm which is additive in repeated bets and to which the law of large numbers applies}" holds true only for lognormal gross returns. The general statement should be rephrased as: \lq \lq \textit{that it is\textbf{ the transformation $\psi^{-1}(\cdot)$ }which is additive and to which the law of large numbers applies}". 

For this reason Carr and Cherubini (2022) call the function $\psi^{-1}(\cdot)$ a logarithmic function in a different algebra, with $\psi(\cdot)$ playing the role of the exponential function. In fact, if one assumes a gamma distributed stochastic clock, leading to the well known Variance Gamma model (first proposed by Madan and Seneta, 1988), this is known as Tsallis algebra in physics.    

Including a stochastic clock in the analysis introduces a source of randomness that operates at the investment horizon $N$ which contrasts the tendency of the law of large numbers to absorb the volatility of the average returns of the investment around the mean. Curiously, using a gamma random variable for the stochastic clock generates a gamma distribution of the average returns which reminds of the model proposed by Browne (1998). \\
Based on (\ref{main}), Carr and Cherubini (2022) proposed to maximize 
\begin{equation}\label{max_CC}
G_N^{CC} = \frac{\psi^{-1}(R_1) + \psi^{-1}(R_2) +\ldots + \psi^{-1}(R_{N-1})+\psi^{-1}( R_N)}{N}
\end{equation}         
where $R_i$ is substituted for $\tilde R_i$, the expected value with respect to the clock. This contains the Kelly-Thorpe rule as a special case when the distribution of the stochastic clock is degenerate:
\begin{equation}\label{max_KT}
G_N^{KT} = \frac{\log(R_1) + \log(R_2) +\ldots + \log(R_{N-1})+\log(R_N)}{N}
\end{equation}         
The present paper addresses the question: what may happen if one manages her bets by taking equation (\ref{max_KT}) instead of (\ref{max_CC})? The framework proposed by Rotando and Thorp 
(1992) provides the ideal environment to carry out this analysis. 

The plan of the paper is as follows. In section \ref{AVG} we derive the average growth rate in a setting with stochastic clocks. In section \ref{Ruin} we extend the definition of ruin as in the Thorp (1969) paper. In section \ref{Model} we discuss how the presence of stochastic clock introduces model risk in long term investment, due to the uncertainty about the probability distribution of the clock. Section \ref{Acceptable} introduces a possible theory of acceptable investments and acceptability indexes to long term investments, based on a set of probability measures of the stochastic clock. Section \ref{Conclusions} concludes.   

\section{Average Growth Rate}\label{AVG}
Consider a sequence of $N$ investment decisions with re-investment of the accrued profits and ask what is the average growth rate of wealth in a setting with time change. The geometrically compounded return is
\begin{equation}
\frac{W_N}{W_0}=\exp\left(\sum_{i=1}^N s_i Z_i\right)
\end{equation}
Define: 
\begin{equation}\nonumber
\bar s \equiv \frac{1}{N}\sum_{i=1}^Ns_i
\end{equation}
the sample average of $s_i$ and:
\begin{equation}\nonumber
\tau_N \equiv \sum_{i=1}^N Z_i
\end{equation}
the cumulated clock. We now may compute the average growth rate as
\begin{eqnarray}
\left(\frac{W_N}{W_0}\right)^{\frac{1}{N}}&=&\exp\left(\sum_{i=1}^N s_i Z_i\right)^{\frac{1}{N}} \nonumber \\
 &=&\exp\left(\sum_{i=1}^N s_i Z_i +\bar s \sum_{i=1}^NZ_i -\bar s \sum_{i=1}^NZ_i -\frac{\tau_N}{N}\sum_{i=1}^N(s_i-\bar s) \right)^{\frac{1}{N}} \nonumber \\
 &=& \exp\left(\frac{1}{N}\sum_{i=1}^N (s_i - \bar s)\left(Z_i-\frac{\tau_N}{N}\right) +\bar s \frac{\tau_N}{N}   \right) \nonumber \\
&=&\exp\left( \text{cov}(s,Z) +\bar s \frac{\tau_N}{N}   \right) \nonumber 
\end{eqnarray}
where cov$(s,Z)$ denotes the sample covariance of $s$ and $Z$. If we assume cov$(s,Z)=0$ we obtain the compounding rule 
\begin{equation}
\frac{W_N}{W_0} =e^{\bar s \tau_N} 
\end{equation} 
If the distribution of the stochastic clock is degenerate, so that $\tau_N = N$ a constant, the average growth rate is turned into a linear average by the logarithmic transformation
\begin{equation}\nonumber
G_N^{KT} \equiv \log \left( \frac{W_N}{W_0}\right)^{\frac{1}{N}} = \bar s = \frac{1}{N}\sum_{i=1}^Ns_i
\end{equation}
Kelly result follows from the fact that the law of large number applies to this linear transformation. If instead $\tau_N=N$ is only a realization of a random variable $\tau(N)$, this transformation is only a possible scenario in an infinite set. Carr and Cherubini (2022) show that computing the average across these scenarios yields a different linear transformation. For the ease of the reader we report the proof here. We denote $E(\cdot)$ the expectation operator with respect to the random variable $\tau(N)$.
\begin{eqnarray}
 E \frac{W_N}{W_0}
 & = &
E  e^{ \bar{s} \tau(N)}
 \nn \\
& = &
E  e^{ \bar{s} \sum_{i=1}^N Z_i}
 \nn \\
& = &
\prod_{i=1}^N
 E e^{\bar{s}  Z_i}
 \mbox{since business time increments are independent }
   \nn \\
   & = &
\prod_{i=1}^N
 E e^{\bar{s}  Z_1}
 \mbox{since business time increments are identically distributed }
   \nn \\
& = &
 \left( E e^{\bar{s} Z_1} \right)^N.
 \label{mtel3}
\end{eqnarray}
Taking the $N$-th root of both sides:
  \be
\left(  E \frac{W_N}{W_0} \right)^{\frac{1}{N}}
= E \left(e^{\bar{s} Z_1}\right)
\equiv  \psi (\bar{s} ),
\label{mtel3a}
\ee
where $\psi$ is the common moment generating function of
the identically distributed business time increments $Z_i$. It follows that the transformation generating the linear average is
\begin{equation}\nonumber
G_N^{CC} \equiv \psi^{-1} \left(\left(E \frac{W_N}{W_0}\right)^{\frac{1}{N}}\right) = \bar s = \frac{1}{N}\sum_{i=1}^Ns_i
\end{equation}
In what follows we will define 
$$G^{KT} \equiv \lim_{N \rightarrow \infty} \log \left( \frac{W_N}{W_0}\right)^{\frac{1}{N}} \quad \quad G^{CC} \equiv \lim_{N \rightarrow \infty} \psi^{-1} \left(\left(E \frac{W_N}{W_0}\right)^{\frac{1}{N}}\right)   $$ 
provided that the limit exists. 
\section{Ruin}\label{Ruin}
It is well known that in the standard Kelly betting problem ruin does not occur. Thorpe (1969) objected that this depends on the definition of ruin probability as $P(W_N =0)=0$. Ruin may occur if one redefines the limit in the almost sure sense. Thorpe (1969) redefined ruin as $P(\lim_{N\rightarrow \infty}W_N < \epsilon)=1 $ for any $\epsilon > 0$. For every investment strategy, ruin is defined as wealth eventually falling short of any positive floor with probability one. In a similar fashion, Stutzer (2010) proposed an investment rule based on shortfall below a target average return. Symmetrically, the same principle can be applied to the definition of \lq \lq favorable game". An investment policy is called \lq \lq favorable" if it eventually guarantees the wealth to grow above any positive threshold $M$: $P(\lim_{N\rightarrow \infty}W_N > M)=1 $.

Assuming that the average growth rate $G$ is a function of a constant investment policy $f$, Thorpe (1969) proves that there exists a boundary $f_c$ between the region of favorable investment and that of ruin. The boundary is $G(f_c)=0$. Strategies $f<f_c$ are in the favorable region, we have $G(f)>0$ and wealth will grow without bounds. For strategies $f>f_c$ we have instead $G(f)<0$ and sooner or later the strategy will run into ruin. The focus of this session, and this paper, is to characterize $f_c$ in the Kelly-Thorpe and the Carr-Cherubini models. There is no need to reformulate Thorp's proof. As a matter of fact the proof only requires the existence of a transformation $\psi^{-1}(\cdot)$ such that the strong law of large numbers apply to the average of the transformed data. Both for the logaritmic function, as in the standard Kelly model, or a moment generating function, as in the Carr-Cherubini model, the proof in Thorp (1969) will carry over in the same way.

We will follow the analysis in Rotando and Thorp (1992) starting from the original Bernoulli game in Kelly (1956). The investment policy is to bet a fixed fraction $f$ of the wealth accumultated at time $t-1$. The outcome of the game will be to have $W_{t-1}(1+f)$ if the bet is successful and $W_{t-1}(1-f)$ otherwise. Denoting $p$ the probability od success and $q\equiv 1- p$ the probability of failure, the function $G^{CC}$ is
$$G^{CC}(f) = p\psi^{-1}(1+f)+q\psi^{-1}(1-f)    $$
The time change model used as reference in this paper is the gamma distributed clock. This leads to the well known Variance Gamma (VG) model, that has replaced the geometric brownian motion model as the main option pricing paradigm. In this case the moment generating function is
$$\psi(s) = (1+\gamma s)^{1/\gamma}  $$
and the transformation function is
$$\psi^{-1}(R) =  \frac{R^\gamma-1}{\gamma} $$
In this case the function $G(f)$ turns out to be
\begin{equation}
G^{VG}(f) = \frac{p(1+f)^\gamma+q(1-f)^\gamma-1}{\gamma}  
\end{equation}
The parameter $\gamma$ is negative: if we denote $\theta$ the variance of the stochastic clock, we set $\gamma = -\theta$.  

\begin{figure}[htbp]
 \centering
 \includegraphics[width=12cm, height=8cm]{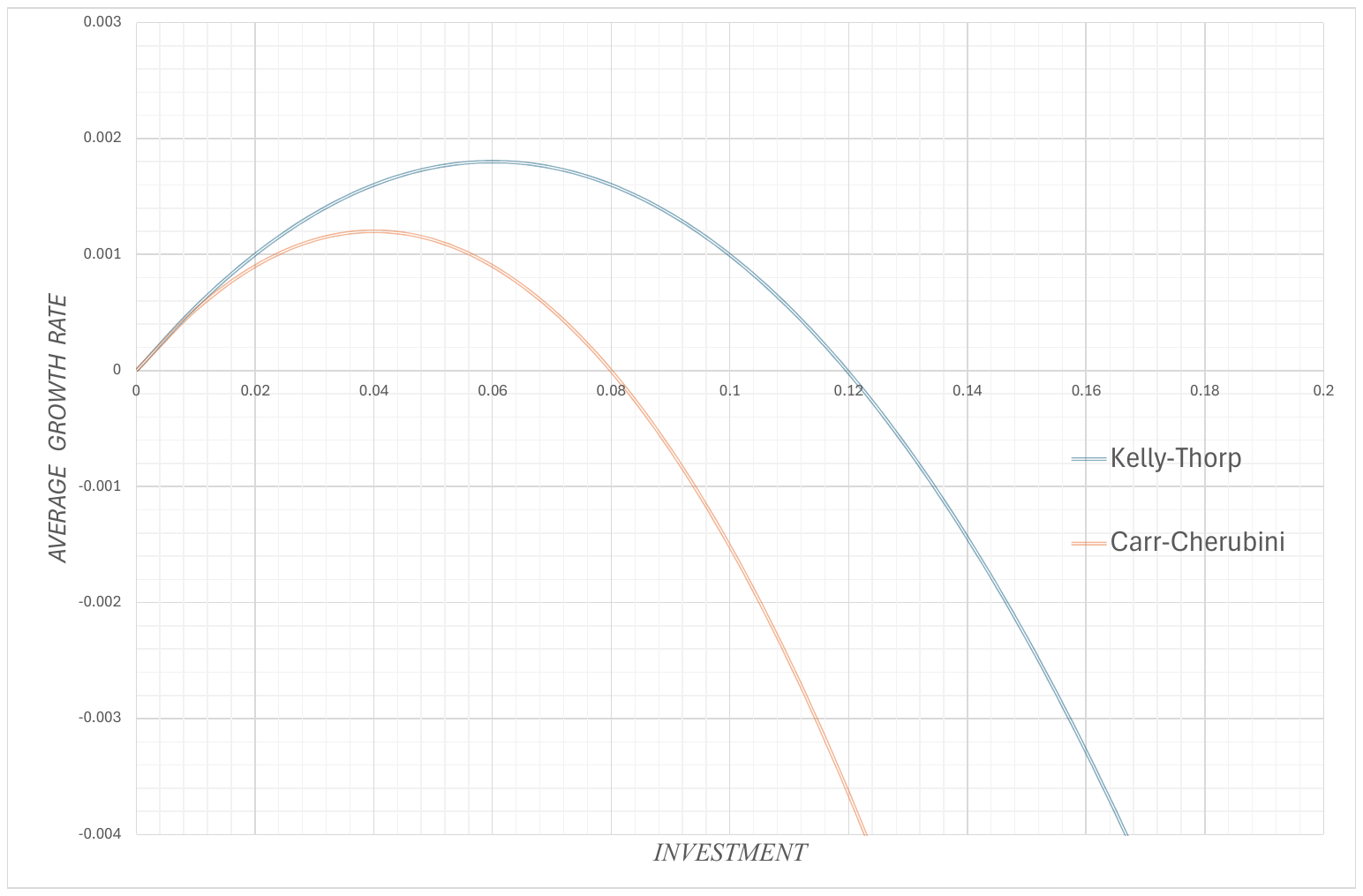} \label{Figure1}
\caption{Average growth rates in the Kelly-Thorpe and the VG and IG models.}
\end{figure}

Figure \ref{Figure1} compares the average growth rates $G^{KT}(f)$ and $G^{VG}(f)$ inspired at Example 1 in Rotando and Thorp (1992), with $p=0.53$. For the VG model, we assume a variance of the clock equal to $0.5$ and we set $\gamma = -0.5$. The figure illustrates quite clearly that if the data generating process of the return is the VG model assumed in the exercise the maximum average growth rate associated to the Kelly rule will never be reached. Moreover, Kelly investment is excessive, meaning that a marginally lower investment would lead to higher average return growth. The ruin threshold, where the average growth function is zero, is also lower in the VG model: we have $f^{VG}_c \approx 0.08$ against $f^{KT}_c \approx 0.12$. However, the Kelly criterion, $f^* = 0.06$, keeps the investor quite safely in the favorable investment region. Curiously enough, only setting $\gamma = -1$ yields $f^{VG}_c=f^*=0.06$, so a Kelly investor would sit exactly on the ruin boundary.  

The same results obtain in the stock market \lq \lq Example 2" discussed in Rotando and Thorp (1992) which assumes a uniform distribution. In the VG model we have 
\begin{equation}\nonumber
G(f)=\int_{LB}^{UB}\frac{(1+fu)^\gamma-1}{\gamma}\frac{1}{UB-LB}du
\end{equation}
To get the maximum, we compute
\begin{eqnarray}
G'(f) &=& \frac{1}{UB-LB}\int_{LB}^{UB}u(1+fu)^{\gamma-1}du \nonumber \\
      &=& \frac{1}{f(UB-LB)} \int_{LB}^{UB}\frac{fs}{1+fu}(1+fu)^{\gamma}du \nonumber \\
      &=& \frac{1}{f(UB-LB)} \left[\int_{LB}^{UB}(1+fu)^{\gamma}du - \int_{LB}^{UB}(1+fu)^{\gamma-1}du \right] \nonumber \\
      &=& 0 \nonumber      
\end{eqnarray}
and we obtain the FOC as
\begin{eqnarray}\nonumber
\frac{1}{\gamma+1}\left[(1+UB f)^{\gamma+1}-(1+LB f)^{\gamma+1} \right]= \frac{1}{\gamma}\left[(1+UB f)^{\gamma}-(1+LB f)^{\gamma} \right]
\end{eqnarray}
The Kelly-Thorp solution is obtained as $\gamma \rightarrow 0$, which yields
\begin{equation}\nonumber
UB-LB = \frac{1}{f}\log \left(\frac{1+UBf}{1+LBf} \right)
\end{equation}
The different in the average growth rates is reported in Figure 3. 
 \begin{figure}[htbp]
 \centering
 \includegraphics[width=10cm, height=8cm]{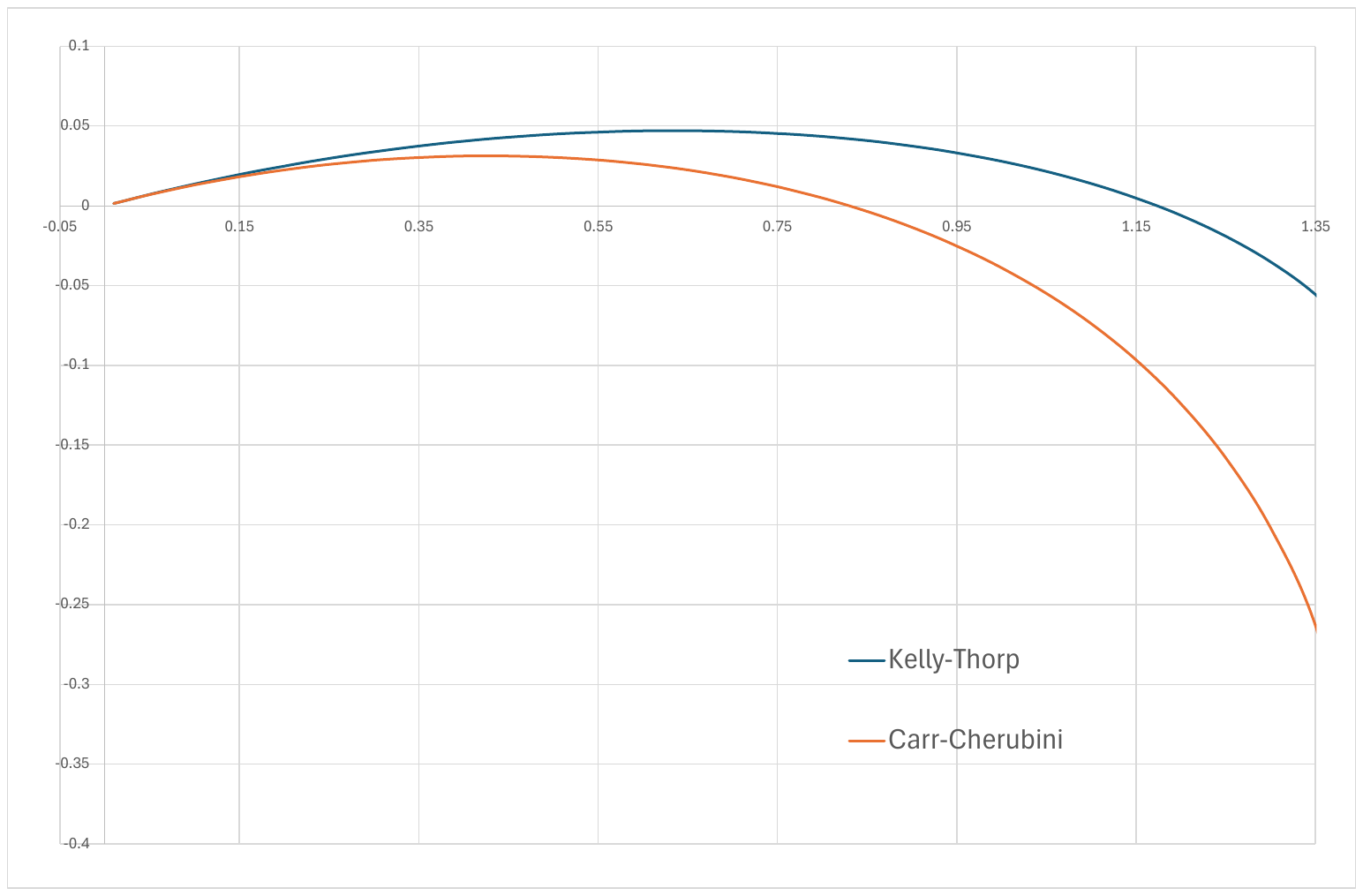} 
\end{figure}

\section{Model Risk}\label{Model}
There is a clear difference between the Kelly-Thorp approach and the Carr-Cherubini one. In the former, the transformation function is the logarithm, and no parameter has to be chosen or estimated. In the latter, a functional form $\psi^{-1}(\cdot)$ must be chosen and a parameter (the variance of the clock) must be estimated. This brings model risk and estimation risk into the picture. While the Kelly-Thorp approach delivers the maximum average growth rate that could be possibly achieved (and will be achieved if the log-returns are normally distributed), the Carr-Cherubini approach suggests that this would not be generally achieved if the log-return follows a general semi-martingale process. But exploiting this information would require to bet on the stochastic process followed by the stochastic clock. 

As for model risk, we provide here the example of the most famous competitor of the VG model, that is the approach in which the clock is assumed to follow a, inverse gaussian (IG) process. This was first proposed by Barndoff-Nielsen (1997). The moment generating function of an inverse gaussian distribution with parameters IG(1,$\lambda$) is
\begin{equation}\label{Laplace_NIG}
\psi(s) = \exp \left(\lambda \left(1- \sqrt{1- \frac{2}{\lambda}s} \right) \right)
\end{equation}
It is matter of little algebra to recover the inverse: 
\begin{equation}
 \psi^{-1}(R) =  \log(R) - \frac{1}{2 \lambda} (\log(R))^2
\end{equation}
The variance of the clock is $\theta = 1/\lambda$. The average growth rate in the binomial case is:
\begin{equation}
G^{IG} = p\log(1+f)+q\log(1-f) - \frac{\theta}{2} \left[(p\log(1+f))^2+q\log(1-f))^2\right]
\end{equation}
Figure 2 depicts the differences between the VG and the IG model, both in comparison with the Kelly-Thorp schedule. In the two models, the variance of the clock is assumed to be the same, namely $\theta = 0.5$. The IG model is much closer to the Kelly-Thorp model, and quite far from the VG one, meaning that the distribution of the stochastic clock may make a substantial difference. 

 \begin{figure}[htbp]
 \centering
 \includegraphics[width=12cm, height=8cm]{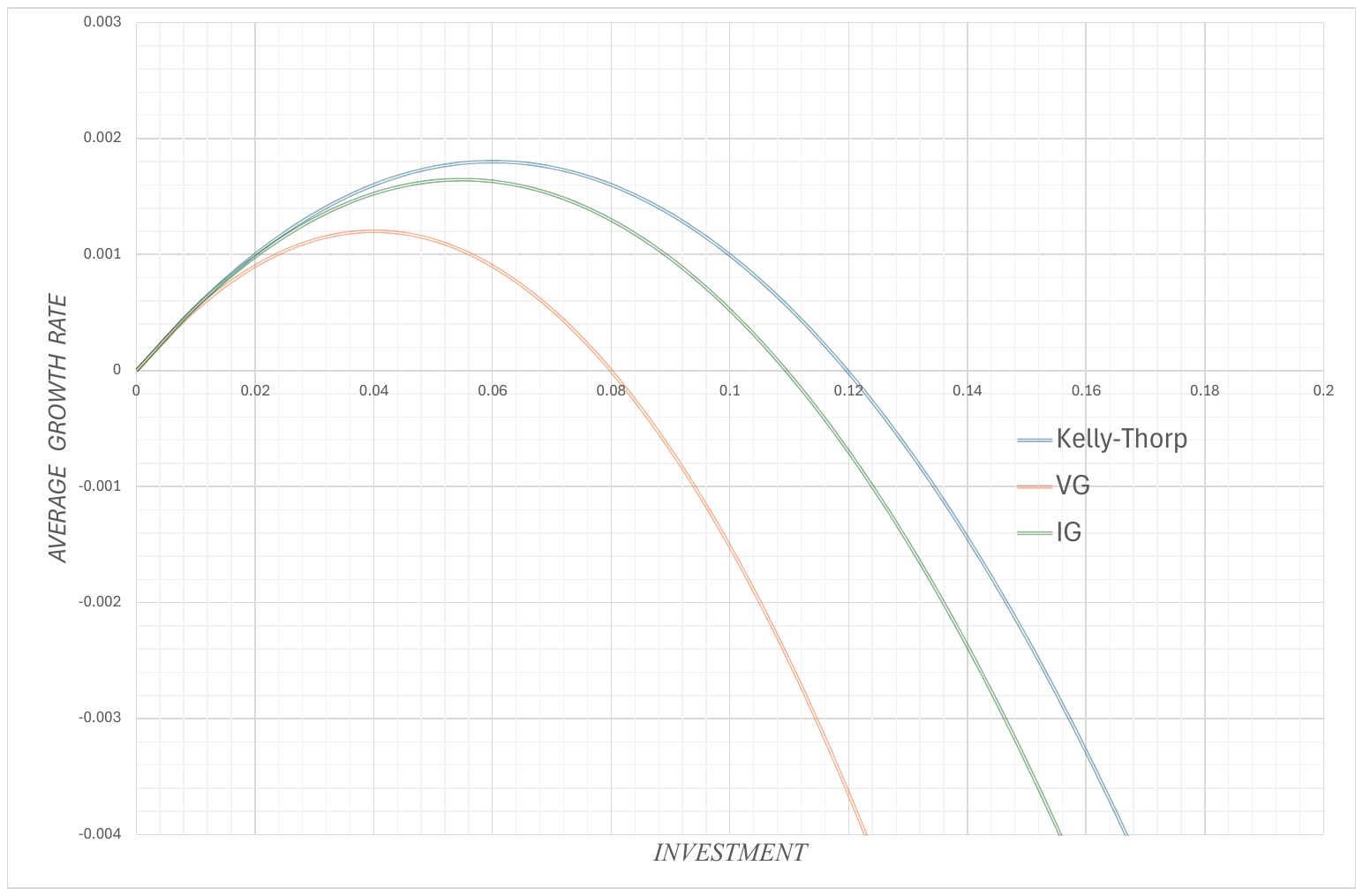} \label{Figure2}
\caption{Average growth rates in the Kelly-Thorpe and the Carr-Cherubini models.}
\end{figure}

Estimation error is also a matter of risk. Table \ref{t1} reports examples of parameter $\theta$ estimates taken from the literature. The estimates are taken from: Hurst et al. (1997) (HPR)\cite{HPR} for the Dow Jones Industrial index , Madan, Carr and Chang (1998) (MCC) \cite{MCC} and Seneta (2004) (SEN)\cite{SEN} for the S\& P 500 on two different samples. Estimates vary from 0.382 to 0.730 for the S\& P 500 market. Also on the same sample two different estimation procedures used in two different papers yield a difference of 0.04 (from 0.382 to 0.422). The presence of outliers may also make a difference. The first two lines of the table show that excluding the 1987 stock market crash decreases the estimate from 0.636 to 0.584. 

The table also reports the effects of model and estimation risk, in the stock market \lq \lq Example 2 " from Rotando and Thorp (1992), whose results are reported in bold in the bottom row. It clearly appears that what really makes a difference is ignoring the stochastic clock, rather than estimating is variance with error. It is striking that whole in the Rotando and Thorp example some leverage (up to 17.1\% ) may be tolerated before entering the ruin region, no leverage at all is allowed if one considers the stochastic clock estimates. Worse yet, investment of the whole wealth is not even allowed, and the investor is at risk of stepping into the ruin region when investment is between 70 and 80\% of wealth. As for the average growth rate actually attained, this would be well below 3\%, against 4.71\% promised under the lognormal return assumption. Moreover, the investment in the stock market is excessive: taking into account the stochastic clock the investment should be decreased from 63.5\% to a range 40-46\%. The maximum rate of growth obtained would then be between 2.87\% and 3.41\%. 

\begin{table}
\begin{center}
\caption{Estimates of the VG model for equity markets }\label{t1}
\begin{tabular}{c|c|c|c|c|c|c}
\hline
\hline
 Market & \small Sample & \small $\theta$ & \small $f_c^{VG}$ & \small $f^*(VG)$ &\small $G^{VG}(f^*(KT)) $ & \small $G^{VG}(f^*(VG)) $  \\
\hline
 \small DJIA HPR\cite{HPR} & \small Jan-73/Dec-93 & 0.636 & 0.7675 & 0.3933 &  0.0173 & 0.0287
 \\ 
  \small DJIA-Ex 1987 & \small Jan-73/Dec-93 & 0.584 & 0.7886 & 0.4061 &  0.0198 & 0.0297 
 \\ 
\hline
 \small S \&P 500 MCC\cite{MCC} & \small Jan-92/Sept-94 & 0.730 & 0.7266 & 0.3720 & 0.0129  & 0.0272 \\
\hline
 \small S \&P 500 SEN\cite{SEN} & \small Jan-77/Dec-81 & 0.422 & 0.8694 & 0.4517 & 0.0274 & 0.0331
 \\
\small  S \&P 500 SEN\cite{SEN} & \small Jan-77/Dec-81 & 0.382 & 0.8941 & 0.4646 & 0.0293 & 0.0341 \\
\hline
%Seneta (2004) &  S \&P 500 & Jan-77/Dec-81 & 0.094353 & 0.1534 & 0.422 & -0.05512 \\
%\hline 
%HPR (1997) & Nikkei & Jan-73/Dec-93  & 0.140076 & 0.156084 & 1.122 & - \\ 
 \hline
    &\small Rotando-Thorp & \textbf{0.000} & \textbf{1.171} & \textbf{0.635} & \textbf{0.0471} \\
 \hline
 \hline
\end{tabular}
\end{center}
\end{table}

\section{Acceptable Long Run Investments}\label{Acceptable}
The presence of model and estimation risk even in long term investment leads to the question of how to handle it and which measures to use to measure the performance. Only taking a conservative estimate of the variance of the clock is only a naif way to approach the problem. Actually, we are in a setting in which uncertainty refers to the set of probability measures with respect to which we compute the performance of the investment. 

It is useful to take the analysis from the repreesentation of the geometric mean in terms of moment generating function, the difference being that the probability measure of the stochastic clock is not known. We only know that it is included in a set $\cal{M}$. 
\be 
\left(  E \frac{W_N}{W_0} \right)^{\frac{1}{N}}
= E_{P \in \cal{M}} \left(e^{\bar{s} Z_1}\right)
\label{mtel3a_multiple}
\ee
We remind the concept of \lq \lq acceptable investments" and  Artzner et al. (1999) and \lq \lq acceptability index" in Madan and Cherny (2009). An index of acceptability of an investment is defined as 
\be \nonumber
\alpha(X) = \sup\{x \in \mathbb{R}_+: \inf_{{Q \in {\cal{D}}}_x} E^Q[X]\geq 0 \} 
\ee
We may extend the same criterion to the average compounded return in equation (\ref{mtel3a_multiple}), reminding that return is positive if it is above 1. We would have
\be \nonumber
\alpha(W) = \sup\left\{x \in \mathbb{R}_+: \inf_{{P \in {\cal{M}}}_x} E^P \left(e^{\bar{s} Z_1}\right)  \geq 1 \right\} 
\ee
Cherny and Madan (2009) suggest a set of acceptability indexes based on the distortion of a proability measure by a function indexed by a positive real number $x$. In this setting the $\alpha(X)$ index of acceptability is given by
\be \nonumber
\alpha(X) = \sup\left\{x \in \mathbb{R}_+:\int_0^\infty y d\left(\phi(F(y) \right)\right\}  
\ee
where $\phi(\cdot)$ is a distortion function, that is an increasing function with the properties: $\phi(0)=0$ and $\phi(1)=1$. In the standard acceptable investment theory this is computed by the Choquet integral. 

The application of the acceptability concept to the long term average return of an investment requires to study the implication of the distortion of a probability measure for the moment generating function. This problem was studied and solved by Mulinacci and Ricci (2025). They define a strictly increasing, continuous and surjective function $g:[a,b]\rightarrow \bar{\mathbb{R}}_+$, with $[a,b]$ a closed subset of $\mathbb{R}$ and $\bar{\mathbb{R}}_+$ the extended interval $[0,\infty]$. In order to use the same function as a distortion function they also impose $g(0)=0$ and $g(1)=1$. The distortion function is then $\phi \equiv g^{-1}$ and they show that if the moment generating function $\psi$ of a probability $P$ is known, and the distortion function $g_x^{-1}$ is applied to it, the pseudo-MGF, that is the MGF associated to the distorted probability distribution is

\be \nonumber
\inf_{{P \in {\cal{M}}}_x} E^P \left(e^{\bar{s} Z_1}\right) = g^{-1}_x(\psi(s))
\ee  

The average growth rate of wealth is then

\be 
 G(f)=\psi^{-1}( g_x(R))
\ee  
 
\section{Conclusions}\label{Conclusions}
In this paper we showed that the statement that maximizing the average log-return of wealth is the best an investor can do in the long run is both true and ambiguous. On one hand it is true that maximizing the log-wealth may lead to the highest possible average return in the long run. On the other hand we have shown that this highest return can be achieved only if the log-return of the investment is normally distributed. If we extend the analysis to the class of processes that may be represented as brownian motions stopped at random time, it turns out that: i) the average growth rate is lower than that predicted by the Kelly rule; ii) the Kelly rule is too aggressive, meaning that investing less in the risky asset would increase the average return; iii) the ruin boundary described by Thorp (1969), with ruin intended as the almost sure probability of the average return falling short of any positive threshold, is lower than in the standard Kelly rule with normal log-returns. 

The result is obtained using the \lq \lq generalized compounding" approach proposed by Carr and Cherubini (2022). The idea is that if one uses the Kelly rule of investment in a general semi-martingale setting, the realized average growth rate of wealth at any future calendar time will be determined by the realization of the business time, that is the stochastic clock, at the same date. The average growth rate of wealth, to which the law of large numbers applies, is then obtained by integrating over the possible realizations of the stochastic clock, and it turns out to be the inverse of the moment generating function of the stochastic clock. In this setting, the average growth rate corresponding the application of the standard Kelly rule will be lower, the higher the variance of the stochastic clock, and the growth rate predicted by the Kelly rule is only obtained as the limit when the variance of the clock approaches zero.  

In this paper we also makes sense of the way in which considering time change can reconcile Kelly and Samuelson, as declared in the title of the Carr and Cherubini (2022), focusing on a key difference between the standard Kelly rule and investment decisions taking into account the stochastic clock dynamics. While the Kelly rule only requires to maximize the log of wealth growth, the maximization proposed by Carr and Cherubini (2022) requires the specification of a model and the estimation of a parameter, the variance of the stochastic clock. This brings back uncertainty into the picture, and in this sense it encounters Samuelson's argument of the relevance of attitude to risk even for long horizon investments. This form of uncertainty persists even if the horizon is infinite, and applies to the argument raised by Ford and Kay (2023). Even an infinitely lived investor would not know the average growth rate she would asymptotically earn, because this is crucially determined by the clock distribution. This brings uncertainty into the picture, even though it is not the kind of uncertainty that is handled under expected utility theory. It is model risk, or knightian uncertainty, rather than a trade off between risk and return. Since Ellsberg (1961) it is well known that expected utility is not well suited to represent this kind of uncertainty. As of today, this problem is handled by the theory of \lq \lq acceptable investments". This is an axiomatic approach based on risk measures, rather than on preferences. This approach, mainly developed by Cherny and Madan (2019), uses distorted probabilities and  Choquet integrals to represent the boundaries of the set of acceptable investments. Since our setting is based on average geometric growth of wealth represented by the moment generating function of the stochastic clock, here we propose to extend the \lq \lq acceptable investment" theory to long run investments exploiting the results in Mulinacci and Ricci (2025) on the moment generating function of a distorted probability measure.

While we hope that this paper clarifies the contribution of the Carr and Cherubini (2022) approach, we think it leaves three important avenues for research open. The first is to further explore measures of performance and risk of long run investments, elaborating and maybe complementing the acceptable investment approach proposed here. The second is to perform empirical work on the actual returns and the relevance of the distribution of the stochastic clock. The third approach is to move beyond the class of processes covered by time change models, that correspond to semi-martingale processes, and to explore other classes that cannot be handled by the time change approach, but that are quite widely used in finance, such as fractionally integrated processes.

\end{document}